\documentclass[reprint,english,aps,pre,floatfix,showpacs]{revtex4-1}

\usepackage{amsmath}
\usepackage{graphicx}
\usepackage{amssymb}
\usepackage{babel}
\usepackage{color}
\usepackage{bm}
\renewcommand{\vec}[1]{\boldsymbol{#1} }

\begin{document}

\title{Effect of controlled  corrugation on capillary condensation of colloid-polymer mixtures}

\author{Andrea Fortini}
\email{andrea.fortini@uni-bayreuth.de}
\author{Matthias Schmidt}

\affiliation{Theoretische Physik II, Physikalisches Institut, Universit\"at Bayreuth, Universit\"atsstra{\ss}e 30, D-95447 Bayreuth,
Germany}

\begin{abstract}
We investigate with Monte Carlo computer simulations the capillary  phase behaviour of  model colloid-polymer mixtures confined between a flat wall and a corrugated wall. The corrugation is modelled via a sine wave as a function of one of the in-plane coordinates leading to a depletion attraction between colloids and the corrugated wall that is curvature dependent.   
We find that for increased amplitude of corrugation the region of the phase diagram where capillary condensation occurs becomes larger. We derive a Kelvin equation for this system and compare its predictions to the simulation results. We find  good agreement between theory and simulation indicating that the primary reason for the stronger capillary condensation  is  an increased contact area between the fluid and the corrugated substrate. On the other hand, the colloid adsorption curves at colloid gas-liquid coexistence show that the increased area is not  solely responsible for the stronger capillary condensation. 
Additionally, we analyse the dimensional crossover from a quasi-2D to a quasi-1D system and find that the transition is characterised by the appearance of a metastable phase. 
\end{abstract}

\maketitle

\section{Introduction}

The equilibrium behaviour of a fluid in contact  with a solid substrate is ruled by the interfacial free energy, i.e., the amount of free energy needed to create the interface~\cite{Rowlinson2002}. 
Since this free energy is the product of the interfacial tension  and the total contact area, the  fluid equilibrium behaviour at a substrate can be controlled by either manipulation of the chemical properties of the surface (interfacial tension) or  the geometry of the surface (contact area).  
It is known that corrugated surfaces have a higher contact area with respect to a flat surface with the same cross-sectional area~\cite{wenzel1936,HAZLETT:1990wh}, and in nature,  corrugation gives the Lotus flower~\cite{Barthlott:1997gn} its characteristic hydrophobicity.  Efforts to mimic the Lotus effect resulted in the production of micromachined surfaces whose wettability has been controlled by proper surface microstructuring~\cite{Onda:1996fb,Patankar:2004hj}.  From a fundamental point of view wetting and capillary condensation on structured~\cite{Dietrich:2005fu} or curved substrates~\cite{Bieker:1998ux,Evans2003} as well as on surfaces with wedge geometry~\cite{Schoen:1997cc,Parry:2000wk} have been studied in detail (see also the review by~\citet{Bonn:2009ha} and references therein). 
Nevertheless, while a lot of research has been done for simple fluids, the effect of corrugation has been neglected in complex colloidal fluids because the roughness of a substrate normally occurs at length scales that are much smaller than the typical size of a colloidal particle. 

Recently, a technique has been introduced that allows the controlled wrinkling of surfaces~\cite{Genzer:2006tu,Schweikart:2009jo} on the micron scale.  The effect of wrinkled substrates on crystallisation has been investigated~\cite{Schweikart:2010kl}, and the resulting structures  have been used to enhance Surface Raman Spectroscopy~\cite{PazosPerez:2010iv}. Clearly, the wrinkling technique could allow a systematic study of the effect of roughness on the phase and  wetting behaviour of complex fluids. 

In this article, we investigate with computer simulations the phase behaviour of a mixture of colloid and non-adsorbing polymers confined between a wrinkled wall and a flat wall.  
We chose colloid-polymer mixtures  because they have been extensively studied in the literature~\cite{Meijer1991,Lekkerkerker1992,Ilett1994,Hoog1999,Dijkstra1999a,Schmidt2000,Brader2000,Tuinier2002a,Aarts2004b,Fortini:2005hq,Vink2005,Aarts2005,Taffs:2010hn}, both from a theoretical and experimental point of view. 
In particular, such mixtures allowed the investigation of a variety of fundamental inhomogeneous statistical phenomena such as wetting at  single flat walls~\cite{Wijting2003a,Wijting2003,Aarts2003,Wessels2005,Brader2003} and capillary condensation and evaporation between flat walls~\cite{Schmidt:2003us, Schmidt:2004he,Fortini:2006fu,Vink2006,Vink2006a,Virgiliis2007}.
Here, we describe the colloid-polymer mixture using the Asakura-Osaawa-Vrij (AOV)~\cite{Asakura:1954jy,Vrij1976} model. Interestingly, the depletion attraction between colloids and the corrugated wall depends on the local curvature of the wall (the same effect occurs for mixtures of hard spheres~\cite{Roth:1999ih,Bryk:2007bl}). 
We find that for increased amplitude of corrugation the capillary condensation is stronger, i.e. the region of the phase diagram (phase space) where capillary condensation occurs becomes larger. 
We also analyse the confined system using a simple thermodynamic description, derive a Kelvin equation and study its validity  by comparing the theoretical predictions to the simulation results. The agreement between theory and simulation is good, and indicate that the increased contact area between the fluid and the substrate (due to corrugation) is the primary cause for the stronger capillary condensation. 
A dramatic increase in colloidal adsorption at the corrugated wall is also primarily due to the increased surface area. On the other hand, further analysis suggests that other effects influence the thermodynamics of the system, like for example the curvature dependence of the particle wall interaction. 
Furthermore, we find that the crossover from a quasi-2D to a quasi-1D system leads to the emergence of metastable phases characterised by a sequence of filled and empty quasi-1D channels.

The paper is organised as follow.
In Sec.~\ref{mod} we introduce the model and the simulation technique. In Sec.~\ref{ke} we derive an expression for the Kelvin equation for corrugated walls, while in Sec.~\ref{res} we present the simulation results. 
Finally in  Sec.~\ref{conc} we draw our conclusions.

\begin{figure*}
\center
\includegraphics[width=12cm]{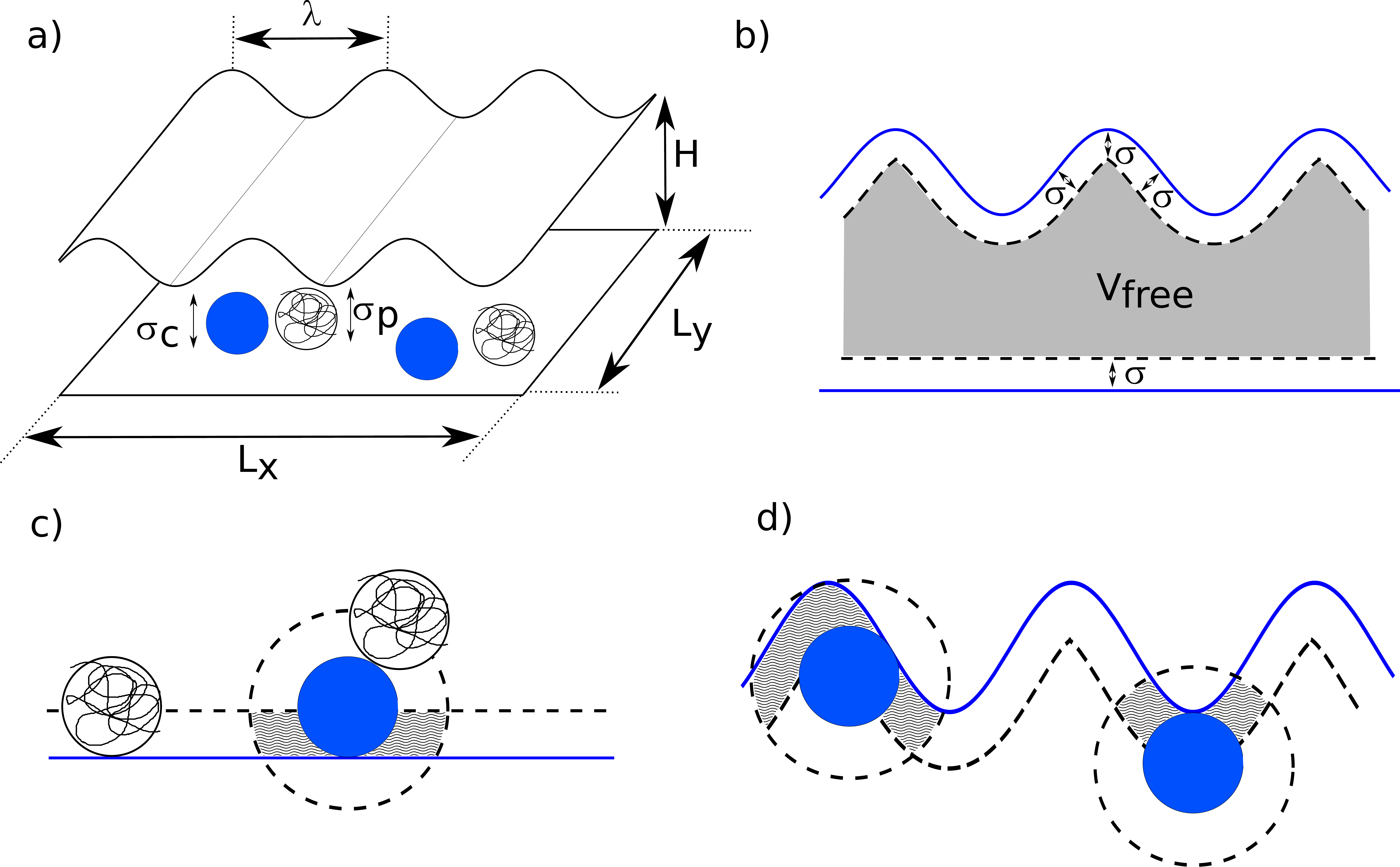}
\caption{a) Schematic drawing of the simulation box and model. Colloids of diameter $\sigma_{c}$ and spheres with diameter $\sigma_{p}$ representing the polymers  are confined between hard walls: a flat wall at the bottom ($z=0$) with area $A_{\rm flat}=L_{x} \times L_{y}$, and a sinusoidal wall of wavelength $\lambda$ and amplitude $a$ at the top ($z=H+a \sin(2\pi x/\lambda)$) .  b) Cross section of the simulation box. The depletion zones  at the walls are delimited by dashed lines. The grey area indicates the volume accessible to the particles. c) Colloids and Polymer at the flat  wall. The depletion zones  are delimited by dashed lines. The grey area indicates the gain in free volume due to the overlap of two depletion regions. d) Like c) but for a colloid at a sinusoidal wall. The gain in free volume depends on the local curvature of the wall. }
\label{fig:model}
\end{figure*}

\section{Model}
\label{mod}

The Asakura-Osaawa-Vrij (AOV)~\cite{Asakura:1954jy,Vrij1976} model is a binary mixture of colloidal hard spheres of diameter $\sigma_c$ and of spheres with diameter $\sigma_p$ representing polymer coils (Fig.~\ref{fig:model}a).
The pair interaction between colloids is that of hard spheres:
$V_{cc}(r)=\infty$ if $r<\sigma_c$ and zero otherwise, where $r$ is
the separation distance between particle centres. The interaction
between a colloid and a polymer is also that of hard spheres:
$V_{cp}(r)=\infty$ if $r<(\sigma_c+\sigma_p)/2$ and zero otherwise. The polymers, however, are assumed to be ideal, hence the polymer-polymer interaction vanishes for all distances, $V_{pp}(r)=0$.
The size ratio $q=\sigma_p/\sigma_c=1$ is a geometric control parameter. 

The colloid-polymer mixture is confined between one smooth planar hard wall at $z=0$ 
and one sinusoidal hard wall at $z=H+a \sin(2\pi x/\lambda)$, where $H$ is a measure of the distance of
the two walls, $a$ is the amplitude of the sinusoidal corrugation,
$\lambda$ is its wavelength, and $x$ is the lateral coordinate perpendicular to the wrinkles.  Any overlap between particles and walls is omitted, therefore the external potentials acting on species $i=c,p$
\begin{equation}
 V_{{\rm ext, }i}(z) = \left\{\begin{tabular}{ll}
 0 & $\vec{r}_i \in V_{\rm free}$\\
 $\infty$ &${\rm otherwise,}$\end{tabular}
 \right.
 \label{EQvext}
\end{equation}
where $V_{\rm free}$ is the free volume available to the particles as sketched in Fig.~\ref{fig:model}b).

The polymers induce an effective colloid-colloid and colloid-wall attraction  that is of entropic origin and is due to the so-called depletion effect. In Fig.~\ref{fig:model}c), we show a colloid in contact with a wall. Around the colloid there is a depletion region (dashed circle) prohibited to the polymers due to the colloid-polymer hard-core interaction (the chains cannot penetrate the colloids). The hard-core interaction between polymers and the wall gives rise to a depletion zone at the wall (dashed line in Fig.~\ref{fig:model}c).
If two colloids approach each other, so that two depletion zones overlap  there is an increase in free volume for the polymer chains, i.e., an increase in entropy \cite{Asakura:1954jy,Vrij1976,Meijer1991,Lekkerkerker1992}. The increase in entropy can be described by an attractive interaction between colloidal particles. 
Likewise, there is an increase in entropy when the depletion zones of colloids and wall overlap (grey zone). The larger the excluded volumes, the stronger is the effective attraction between the hard wall and the colloids. 
 Figure~\ref{fig:model}d) illustrates the depletion zone of a curved wall. Clearly, the gained free volume (grey zones)  depends on the local curvature of the wall. The concave part of the wall leads to a larger gain in free volume (more attractive) than the convex part of the wall (less attractive).

We denote the packing fractions by $\eta_i = \pi\sigma_i^3 N_i/(6AH)$,
where $N_i$ is the number of particles of species $i$ and $A$ is the
cross-sectional (normal to the $z$-direction) area of the wrinkled wall, and also equal to the area of the flat wall. 
Additional thermodynamic quantities are the scaled chemical potential $\beta \mu_c$ and $\beta \mu_p$ of colloids and polymers, respectively, and the polymer reservoir packing fraction $\eta_{p}^{r}=\exp(\beta \mu_p)$ of a reservoir of pure polymers that is in chemical equilibrium with the system. 
The inverse temperature is $\beta=1/k_{B} T$, with $k_{B}$ the Boltzmann constant and $T$ the temperature

\section{Simulation method}
Fig.~\ref{fig:model}a) displays an illustration
of the model, for which we carried out Monte Carlo computer
simulations for a wide range of different values of particle
concentrations and for several values of the amplitude $a$. The
wavelength of the corrugation was fixed to the value of
$\lambda=10 \sigma_c$.  Periodic boundary conditions were applied in the
$x$- and $y$-directions, and the box size in the
$x$-direction was chosen as $4\lambda$.

We carried out Monte Carlo simulations in the grand canonical
ensemble, i.e. with fixed volume, temperature, and chemical
potentials $\beta \mu_c$ and $\beta \mu_p$ . 

To study the phase coexistence, we sample the probability
$P(N_c)|_{ \mu_c, \mu_p}$ of observing $N_c$ colloids in a volume
$V$ using the successive umbrella sampling~\cite{Virnau2004}. 
We use the histogram reweighing technique to obtain the probability distribution for any $ \mu_c'$
once $P(N_c)|_{ \mu_c, \mu_p}$ is known for a given $ \mu_c$:
\begin{equation}
\ln P(N_c)|_{ \mu_c', \mu_p} = \ln P(N_c)|_{ \mu_c, \mu_p} +
N_c (\beta \mu_{c}'-\beta \mu_{c}) 
\label{histogram}.
\end{equation}
At phase coexistence, the distribution function $P(N_c)$ becomes
bimodal with two separate peaks of equal area for the colloidal
liquid and gas phases.  We determine which $\mu_c'$ satisfies the equal area rule
\begin{equation}
\int_{0}^{\langle N_c \rangle }P(N_c)|_{ \mu_c', \mu_p}dN_c =
\int_{\langle N_c \rangle }^{\infty} P(N_c)|_{ \mu_c', \mu_p}dN_c,
\end{equation}
with the average number of colloids
\begin{equation}
\langle N_c \rangle=
\int_{0}^{\infty}N_cP(N_c)|_{ \mu_c, \mu_p}dN_c,
\end{equation}
using the histogram reweighing equation (\ref{histogram}). 
The sampling of the probability ratio $P(N)/P(N+1)$ is done, in each window, until the difference between two
successive samplings of the probability ratio is smaller than $1 \times 10^{-4}$.
To improve the sampling accuracy, we used the cluster move introduced by~\citet{Vink2004d}.

\section{Theory: Kelvin Equation}
\label{ke}
We apply a thermodynamic treatment in the limits of $\lambda/\sigma \gg 1$ and $H/\sigma \gg1$ to the system sketched in Fig.~\ref{fig:model}. 
The derivation follows closely the derivation of~\citet{Evans:1990bx} for fluids between smooth parallel walls. 
For simplicity we limit the theoretical derivation to the case of one-component fluids. As will become clear at the end the section, the result can be generalised effortlessly to multi-components fluids. 

The free energy in the grand canonical ensemble is the grand potential $\Omega$. For a fluid between a top and a bottom walls of area $A_{1}$ and $A_{2}$, respectively, the grand potential is the sum of the bulk free energy per unit volume $\omega$ and the surface energy contributions from the two walls,
\begin{equation}
\Omega(\mu)=V \omega(\mu) + A_{1} \gamma^{1}(\mu)+A_{2}\gamma^{2}(\mu) \ ,
\label{omega}
\end{equation}
where $V$ is the volume, $\mu$ is the chemical potential and $\gamma^{1}, \gamma^{2}$, are the fluid-wall interfacial tensions for the top and bottom wall, respectively.

We introduce the bulk chemical potential at coexistence $\mu_{b}$, and use it as a reference state for the chemical potential of the confined system
$$\mu=\mu_{b}+\Delta \mu.$$ Assuming $\Delta \mu$  small we Taylor expand equation (\ref{omega}) around $\mu_{b}$, recalling that the bulk density is $$\rho=-\frac{\partial \omega}{\partial \mu}$$ and that the adsorption is $$\Gamma=\frac{\partial \gamma}{\partial \mu}.$$
The Taylor expansion yields
\begin{eqnarray}
\frac{\Omega(\mu)}{V}&=& \omega(\mu_{b})-\rho(\mu_{b}) \Delta \mu + \frac{A_{1}}{V} \gamma^{1}(\mu_{b}) \nonumber \\&+&\frac{A_{1}}{V} \Gamma^{1}(\mu_{b}) \Delta \mu +\frac{A_{2}}{V}\gamma^{2}(\mu_{b})+\frac{A_{2}}{V}\Gamma^{2}(\mu_{b}) \Delta \mu.
\label{omegaT}
\end{eqnarray}
The coexistence between a liquid and a gas phase inside the slit occurs when both thermal and mechanical equilibrium conditions are satisfied, i.e., the chemical potential and pressure of the gas and liquid phases are the same. These two conditions are satisfied when the grand potential of the two phases is the same 
$\Omega^{\rm gas}(\mu)=\Omega^{\rm liq}(\mu)$.
Given that  $\omega^{\rm liq}(\mu_{b})= \omega^{\rm gas}(\mu_{b})$, the above equilibrium condition inside the slit leads to the following relation 
\begin{eqnarray}
0&=&( \rho_{\rm liq}- \rho_{\rm gas}) \Delta \mu + \nonumber \\&+& \frac{A_{1}}{V} (\gamma^{1}_{\rm gas}-\gamma^{1}_{\rm liq}) +\frac{A_{1}}{V}(\Gamma^{1}_{\rm gas}- \Gamma^{1}_{\rm liq})\Delta \mu + \nonumber \\&+&\frac{A_{2}}{V}(\gamma^{2}_{\rm gas}-\gamma^{2}_{\rm liq}) +\frac{A_{2}}{V}(\Gamma^{2}_{\rm gas}- \Gamma^{2}_{\rm liq})\Delta \mu \ .
\label{K1}
\end{eqnarray}
We stress that the above relation (\ref{K1}) is valid for any type of surface, as we only requested that $\Delta \mu$ be small.

The previous relation greatly simplifies  if we further assume that the adsorption at the wall is negligible and that $\gamma^{1}=\gamma^{2}=\gamma$, in which case
\begin{eqnarray}
 \Delta \mu= (\frac{A_{1}}{V}+\frac{A_{2}}{V})  \frac{(\gamma_{\rm liq}-\gamma_{\rm gas})}{( \rho_{\rm liq}- \rho_{\rm gas})} \label{K2}
\end{eqnarray}

We stress that the assumption $\gamma^{1}=\gamma^{2}=\gamma$ implicitly considers a curvature independent interfacial tension. That is, we  treat the limit of $a/\sigma \gg 0$. This apparently crude approximation is justified in the framework of the Kelvin equation that treats systems in the limit $H/\sigma \gg1$.
Strictly speaking, the Kelvin equation is not valid for strongly confined systems, but has been shown before that its predictions remain qualitatively good down to quasi-2D systems. This theoretical description would on the other hand  break down in the quasi-1D limit, because it would predict a phase transition that in reality does not exists for a bulk one-dimensional system.

\subsection{Two flat walls}
If we  consider two flat walls of equal area $A_{1}=A_{2}=A^{\rm flat}$, and volume $V=A^{\rm flat} H$ we obtain the standard equation~\cite{Evans:1990bx} 
\begin{eqnarray}
 \Delta \mu^{\rm flat}= \frac{2}{H}  \frac{(\gamma_{\rm liq}-\gamma_{\rm gas})}{( \rho_{\rm liq}- \rho_{\rm gas})} .
 \label{Kf}
\end{eqnarray}

\begin{figure}
\includegraphics[width=8.5cm]{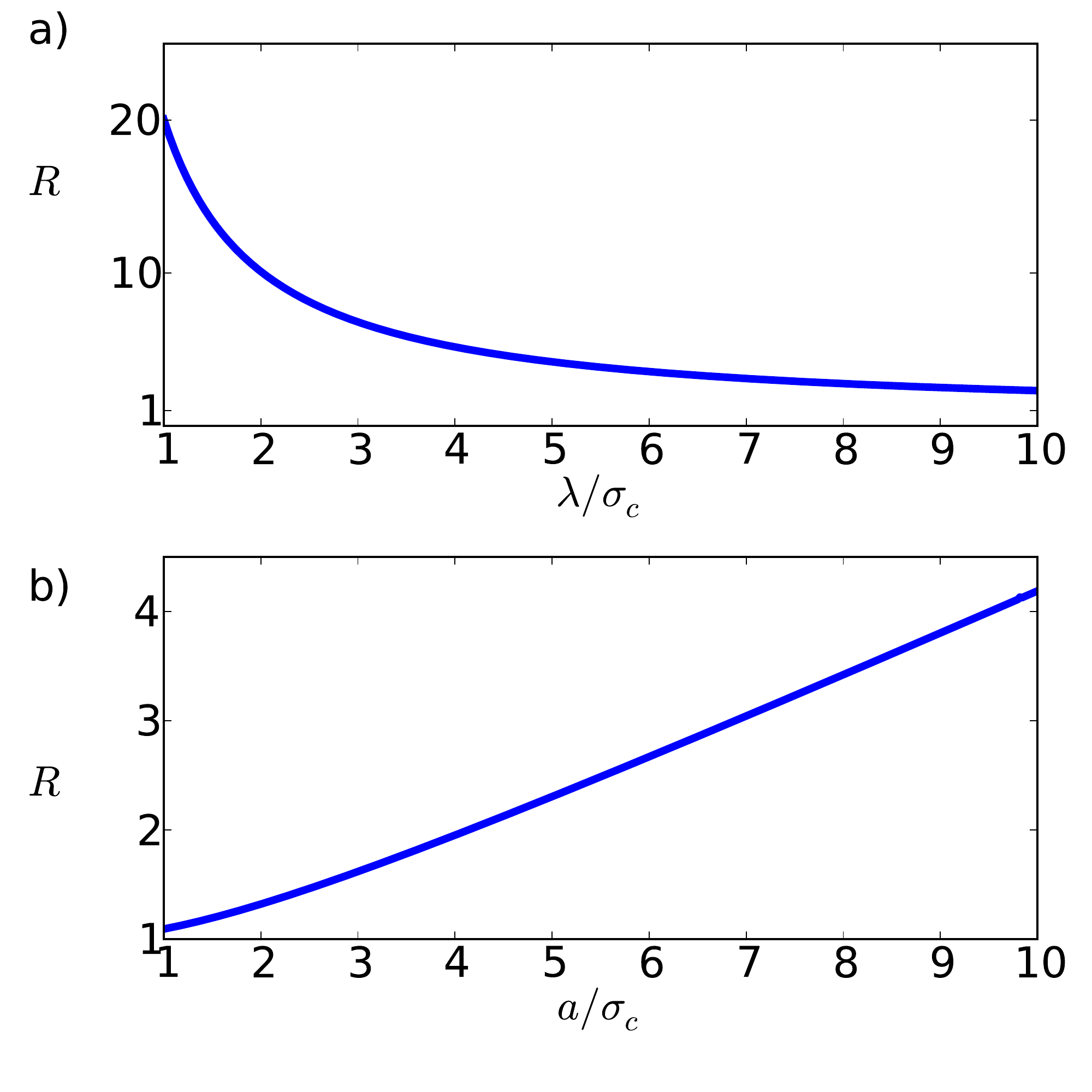}
\caption{The roughness ratio $R$ for a sinusoidal wall. a) For fixed amplitude $a=5 \sigma_{c}$ and changing wavelength. b) For wavelength $\lambda=10 \sigma_{c}$ and changing amplitude.}
\label{R}
\end{figure}

\subsection{One corrugated wall and one flat wall}
A system with one flat wall and one corrugated wall has a  volume $V=A^{\rm flat} \langle H \rangle$, with $ \langle H \rangle$ the average separation distance between the walls. 
The relation (\ref{K2}) then becomes
\begin{eqnarray}
 \Delta \mu^{\rm corr}= \frac{1}{\langle H \rangle} (1+R)  \frac{(\gamma_{\rm liq}-\gamma_{\rm gas})}{( \rho_{\rm liq}- \rho_{\rm gas})} ,
 \label{K3}
\end{eqnarray}
where $$ R=\frac{A_{\rm corr}}{A_{\rm flat}}$$ is the ratio between the flat and corrugated area and  is equivalent to the Wenzel roughness ratio defined as the ratio between the contact area and the geometric cross sectional area~\cite{wenzel1936}.

By comparing Eqs. (\ref{Kf})  and (\ref{K3})  we find also that the ratio between the chemical potential shifts of the corrugated and flat walls is related to the Wenzel ratio via 
\begin{equation}
 \Delta \mu^{\rm corr}/ \Delta \mu^{\rm flat} = \frac{1+R}{2} \ .
 \label{Kr1}
\end{equation}

\subsection{Two corrugated walls}
For the case of two corrugated walls the  general Kelvin equation (\ref{K2}) becomes

\begin{eqnarray}
 \Delta \mu^{\rm corr}= \frac{2}{\langle H \rangle} R  \frac{(\gamma_{\rm liq}-\gamma_{\rm gas})}{( \rho_{\rm liq}- \rho_{\rm gas})} ,
 \label{K4}
\end{eqnarray}
By comparing Eqs. (\ref{K4}) and (\ref{Kf}) we find that for the case of two corrugated walls  the ratio is  
$$  \Delta \mu^{\rm corr}/ \Delta \mu^{\rm flat} = R.$$

\subsection{Generalisation to binary mixtures}
The derivation of the Kelvin equation for binary mixtures~\cite{Evans:1987ud} proceeds along the lines described above for a one-component fluid. A closed set of equation for the chemical potential shifts $\Delta \mu_{c}$, and  $\Delta \mu_{p}$, for colloid and polymer, respectively, can however be obtained only when an independent relationship between $\Delta \mu_{c}$, and  $\Delta \mu_{p}$ is used. In Ref.~\citenum{Fortini:2006fu} three possible choices for the relationship are  outlined.  Independent of the choice of the relationship, the equations for binary mixtures for corrugated walls leads to
\begin{eqnarray}
 \Delta \mu_{c}^{\rm corr}/ \Delta \mu_{c}^{\rm flat} &=& \frac{1+R}{2} \\ \nonumber
  \Delta \mu_{p}^{\rm corr}/ \Delta \mu_{p}^{\rm flat} &=&  \frac{1+R}{2} \ ,
 \label{Kb2}
 \end{eqnarray}

 and
\begin{eqnarray}
 \Delta \mu_{c}^{\rm corr}/ \Delta \mu_{c}^{\rm flat} &=& R   \\ \nonumber
  \Delta \mu_{p}^{\rm corr}/ \Delta \mu_{p}^{\rm flat} &=& R  \ , 
 \label{Kb1}
  \end{eqnarray}
for one and two corrugated walls, respectively.

\subsection{Roughness ratio for a sinusoidal wall}
We consider a wall corrugated in one direction by a sinusoidal functional form as shown in the sketch of Fig.~\ref{fig:model}a). 
The cross sectional area is $A_{\rm flat}=L_{x} \times L_{y}$, where $L_{x}$ and $ L_{y}$ are the parallel and perpendicular directions  with respect to the sinusoidal direction, and $L_{x}=n \lambda$, where $n$ is an arbitrary number and $\lambda$ is the wavelength.
Therefore, the roughness ratio can be written as 
\begin{equation}
R=\frac{\int_{0}^{n \lambda} dx \sqrt{1 +   \frac{4 \pi^{2}}{\lambda^{2}} \cos^{2}(\frac{2 \pi}{\lambda} x  )  } }{n \lambda} \ ,
\label{eq:r}
\end{equation}
where  the numerator is the line integral over the sinusoidal path. The elliptical integral can be solved numerically. The value of the integral (\ref{eq:r}) is the same for any integer value of $n$, but changes when non-integer values of $n$ are chosen. Therefore, an explicit dependence of the integral on $n$ is left for the general case.

Figure~\ref{R} shows the behaviour of the roughness ratio defined in Eq.~(\ref{eq:r}).
For fixed amplitude and increasing wave length (Fig.~\ref{R}a) the ratio $R$ decreases. Clearly for $\lambda \rightarrow \infty$ the ratio goes to one.
On the other hand, for fixed wavelength $\lambda$ (Fig.~\ref{R}b) the roughness ratio $R$ increases monotonically for increasing values of the amplitude of corrugation $a$.

\section{Simulation Results}
\label{res}

\subsection{Phase diagram}
\begin{figure}
\includegraphics[width=8.5cm]{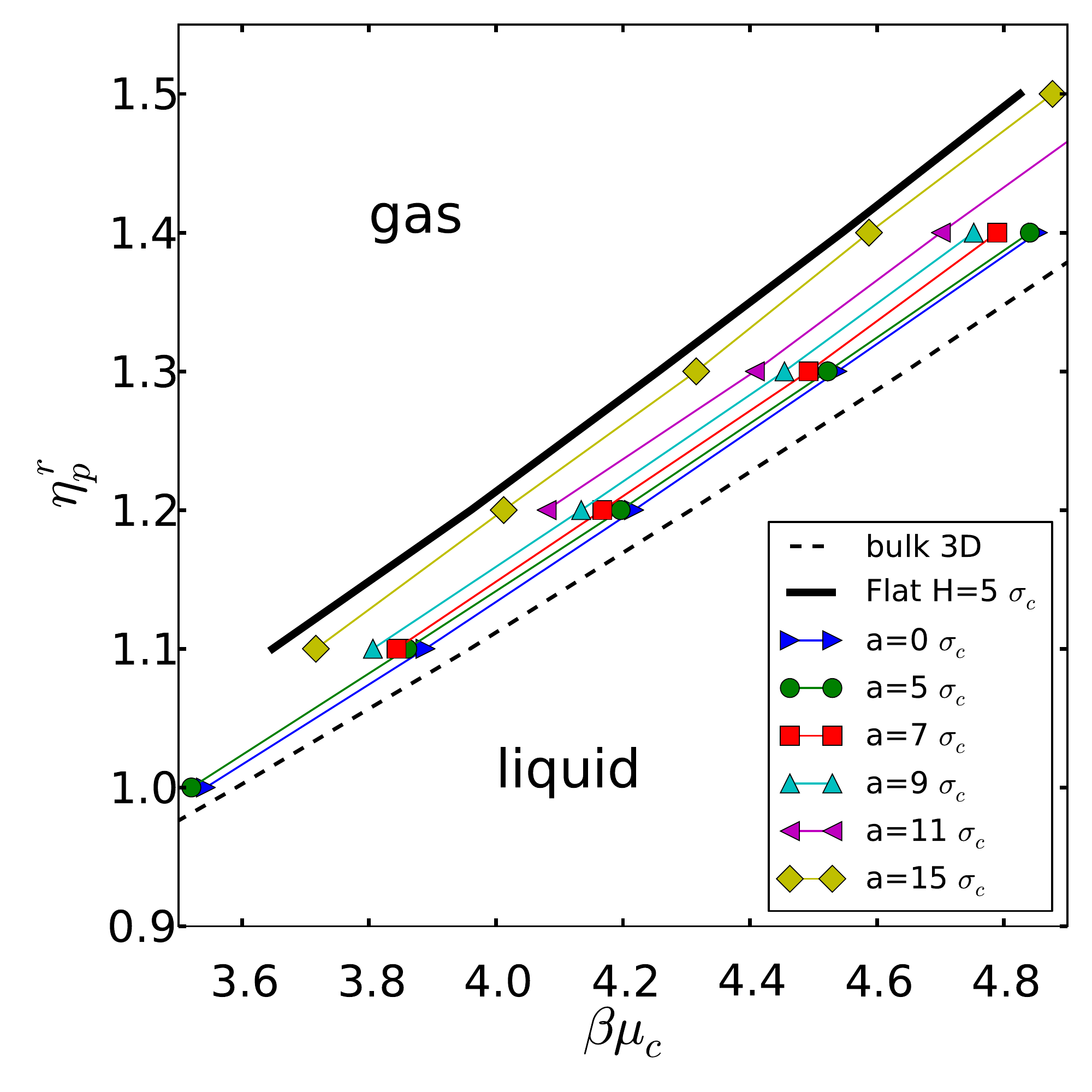}
\caption{Phase diagram in the  ($\beta \mu_{p},\beta \mu_{c}$) representation. Shown are the binodals for wavelength $\lambda=10 \sigma_{c}$ and  amplitudes  $a/\sigma_{c}$=0, 5, 7, 9, 11, 13, and 15. Also shown are the binodals for the bulk system (dashed line) and for the system confined between two flat walls at distance $H=5 \sigma_{c}$ (thick continuos line).}
\label{fig:phd}
\end{figure}
Figure~\ref{fig:phd} shows the phase diagram in the ($ \beta \mu_{p}$, $\beta  \mu_{c}$) representation. Each line represent a binodal, i.e. the value of colloid and polymer chemical potentials at which phase separations occurs. For chemical potentials in the region above a binodal the gas phase is stable, whereas in the region below the binodal the liquid phase is stable.  In particular, the capillary binodals for corrugation amplitude $a/\sigma_c=0-15$ are shown (symbols). As reference we also show  the bulk binodal (dashed line), and the capillary binodal for flat walls with separation distance $H/\sigma_c=5$ (thick continuous line).
The  binodal for the colloid-polymer mixtures confined between two flat walls (i.e., $a$ = 0  $H/\sigma_c$ = 15) is shifted toward smaller colloidal chemical potentials and higher polymer chemical potential with respect to the bulk binodal (thin dashed line). The shift of the binodals indicate the occurrence of capillary condensation, because in the region between the confined and bulk binodals,  the stable phase for the confined system is the liquid, while for the bulk system is the gas.  This effect  was extensively studied in the literature~\cite{Schmidt:2003us,Fortini:2006fu,Vink2006,Fortini:2008ha} for colloid-polymer mixtures confined between flat walls. 

Here, for  colloid-polymer mixtures confined between one flat and one corrugated wall, we find that for increasing corrugation, i.e., increasing  amplitude $a$ of the sinusoidal wall, the binodals shifts toward larger polymer chemical potentials and smaller colloid chemical potentials, indicating progressively stronger capillary condensation with respect to the flat walls, i.e., larger region of phase space where the capillary condensation occurs. 
The critical points were not calculated because an accurate determination would require a careful finite-size analysis~\cite{Vink2006} that is beyond the scope of the current work. The end-points in Fig.~\ref{fig:phd} were determined as the lowest value of the chemical potentials at which a double peak in the probability distribution was observed. 

One trivial interpretation for the behaviour is that the sinusoidal wall introduces a length scale in the system that is smaller than the average wall separation distance, namely $h_{min}=H-a$, that strengthens the capillary condensation.  However, this interpretation is not supported by our results. 
For example for $a = 11 \sigma_{c}$, the smallest length scale is $h_{min}=15\sigma_{c}-11\sigma_{c}=4 \sigma_{c}$. 
As a reference in Fig.~\ref{fig:phd} we plot the binodal for two flat walls at distance $H/\sigma_{c}=5$. The $a =11 \sigma_{c}$ binodal (left triangles) is clearly separated from it.

For amplitude $a=15\sigma_{c}$ the system consist of a series of independent quasi-1D channels. 
Interestingly, we find that the binodal lines for increasing amplitudes $a$, slowly approach the binodal of the system with a single groove, therefore the system undergoes a slow dimensional crossover from a quasi-2D to a quasi-1D system.  
We stress that in one-dimensional systems the gas-liquid phase transition does not exist in the thermodynamic limit. Even in quasi-1D pores a rounding of the transition is noted~\cite{Wilms:2010gx} and a multi domain structure is observed instead of a proper gas-liquid separation. 

\subsection{Comparison with the Kelvin equation}
The Kelvin equation~(\ref{Kr1}) suggests, on the other hand that the stronger capillary condensation for increasing amplitudes $a$ is due to the increased surface area of the corrugated wall. To evaluate the validity of this interpretation we compare theoretical results with the simulation findings. In particular, we compare the ratio $\Delta \mu_{c}^{\rm corr}/\Delta \mu_{c}^{\rm flat}$ found in simulations to the theoretical result of Eq.~(\ref{Kr1}). 
Remarkably, the simulation results start at smaller values of the ratio but they follow the same trend as the theory results for increasing amplitudes $a/\sigma_{c}$, demonstrating that the stronger capillary condensation is mainly driven by the increased surface area of the corrugated wall. 
We believe that the smaller values of the ratio $\Delta \mu_{c}^{\rm corr}/\Delta \mu_{c}^{\rm flat}$ in simulations with respect to the theory are due to the large adsorption of colloids at the sinusoidal wall in the gas phase. 
The Kelvin equation also neglects the curvature. Recent works on curved surfaces~\cite{Troster:2012cz,Laird2013} suggest that the inclusion of curvature effects  would lead to  smaller values of the $\gamma_{\rm liq}-\gamma_{\rm gas}$ difference. This leads to a decrease of the shift in chemical potential predicted by the Kelvin equation, leading to a better agreement with simulation results.  

\begin{figure}
\includegraphics[width=8.5cm]{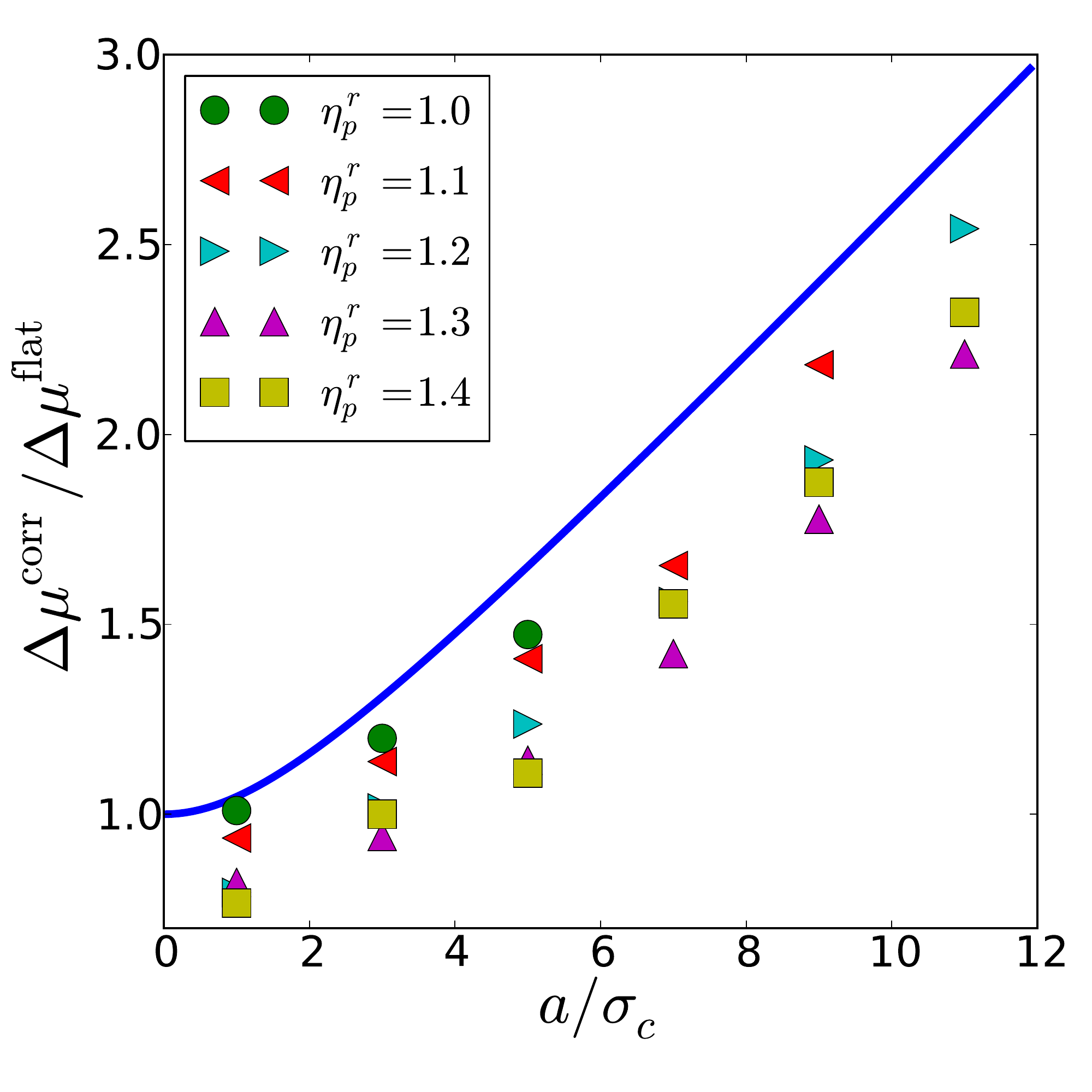}
\caption{The ratio $\Delta \mu_{c}^{\rm corr}/\Delta \mu_{c}^{\rm flat}$ for different amplitudes $a$ of the sinusoidal wall at fixed wavelength $\lambda=10\sigma_{c}$. We compare results of the computer simulations (symbols) for different values of polymer reservoir packing fraction $\eta_{p}^r$ with the theoretical prediction (line) of  equation (\ref{K3}). }
\label{cmp}
\end{figure}

\subsection{Adsorption}

\begin{figure}
\includegraphics[width=8.5cm]{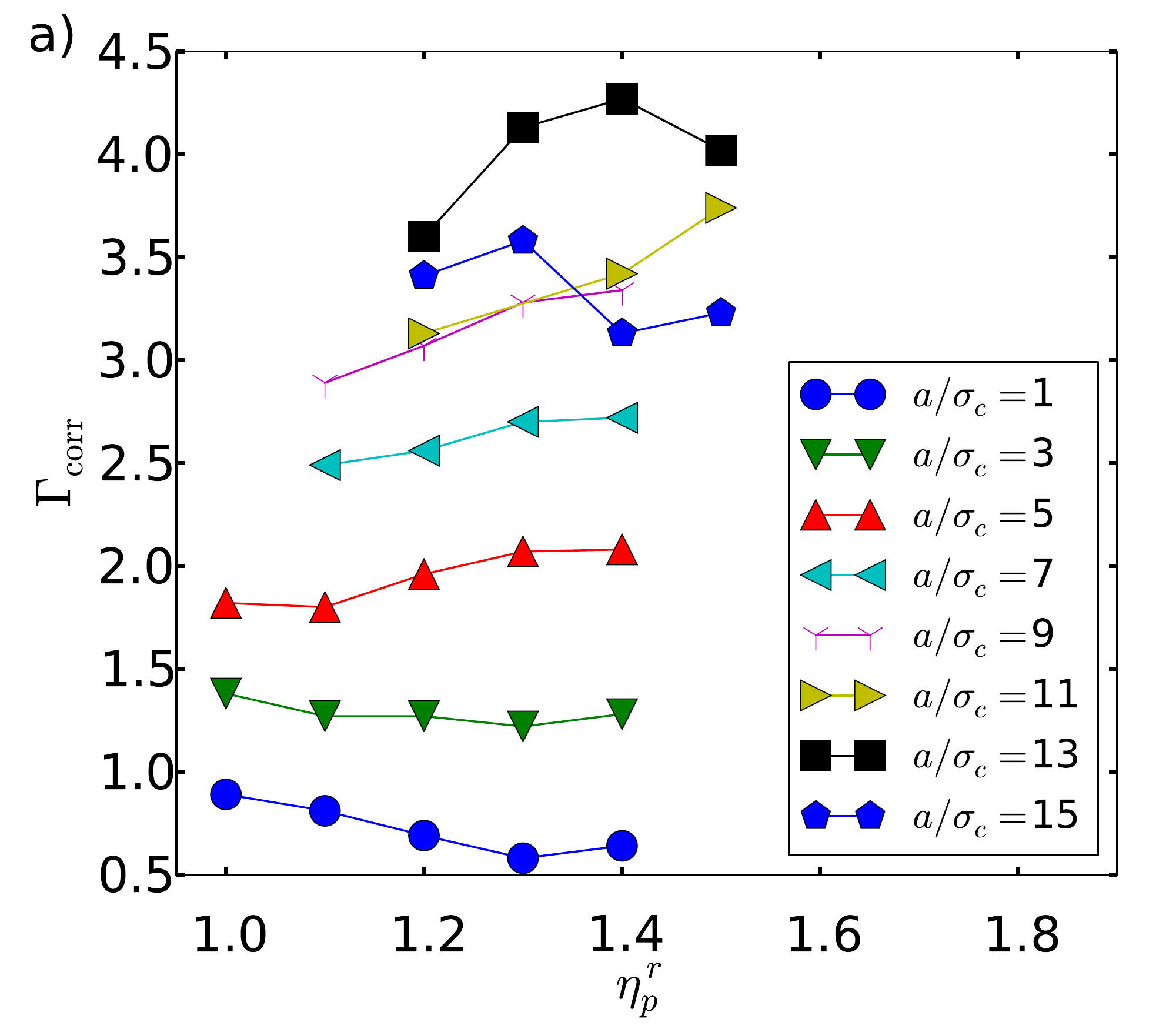}
\includegraphics[width=8.5cm]{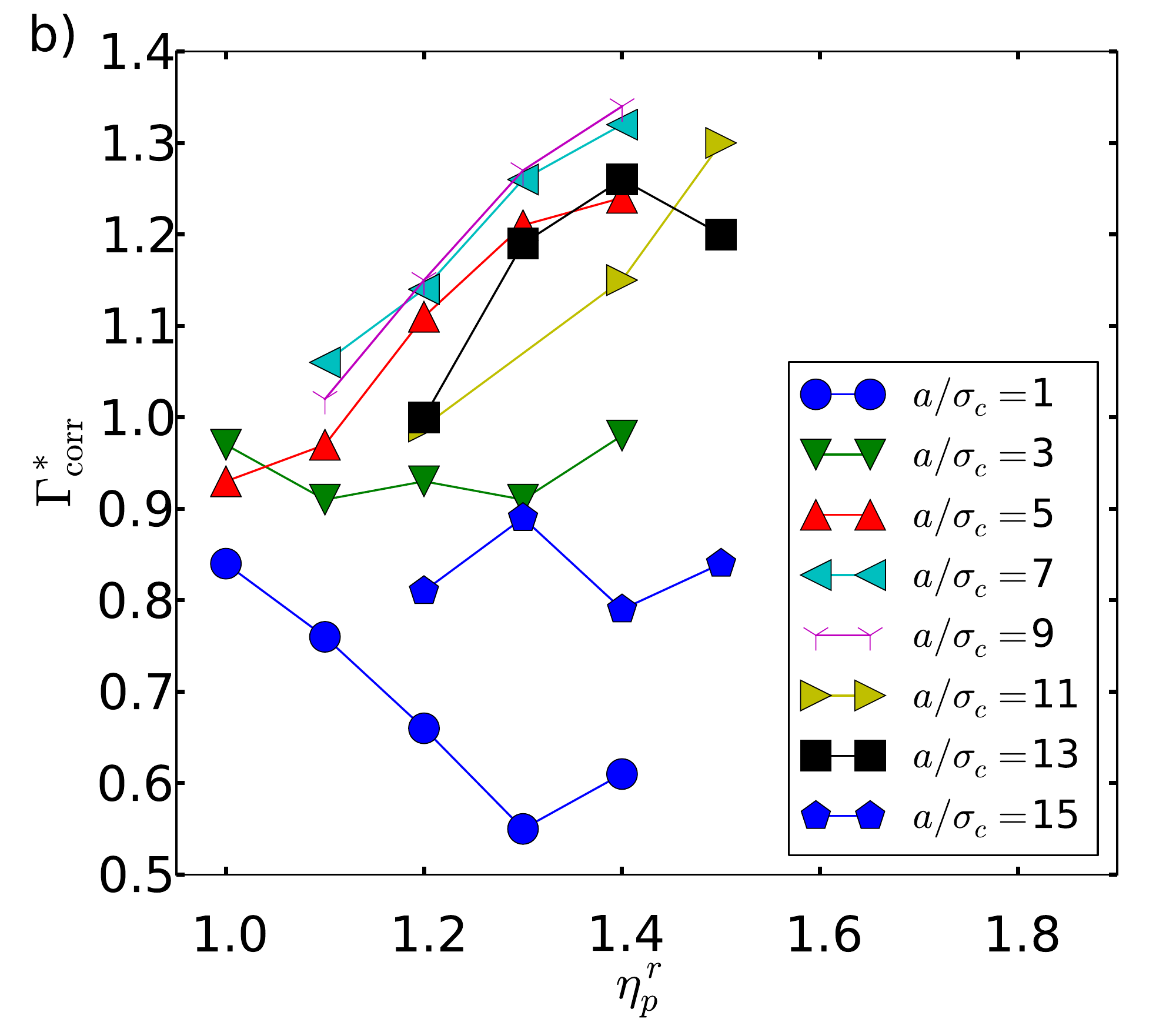}
\caption{Adsorption  at the corrugated wall for increasing values of the corrugation amplitude $a/\sigma_c$. a) Colloidal adsorption $\Gamma_{\rm corr}$ with the cross-sectional area  $A^{\rm flat}$ as reference surface. (Eq.~(\ref{eq:adfl})). b) Colloidal adsorption $\Gamma^*_{\rm corr}$ with the real contact area  $A^{\rm corr}$ as reference surface (Eq.~(\ref{eq:adcr})).}
\label{fig:ads}
\end{figure}

Further insights can be gained by calculating the colloid adsorption $\Gamma$ for state points at gas coexistence.
First of all, we carried out simulations of colloid-polymer mixtures between two flat walls at separation $H=15 \sigma_{c}$ to measure the adsorption at one flat wall 
\begin{equation}
\Gamma_{\rm flat}=\frac{N_{c}^{\rm flat}-N_{c}^{\rm bulk}}{2 A^{\rm flat}}
\end{equation}
at the gas coexistence state points of the phase diagram shown in Fig.~\ref{fig:phd}.
Subsequently, we performed simulations for the mixtures confined between one flat and one sinusoidal wall and computed the adsorption at a single sinusoidal wall 
\begin{equation}
\Gamma_{\rm corr}=\frac{N_{c}^{\rm corr}-N_{c}^{\rm bulk}}{A^{\rm flat}}-\Gamma_{\rm flat} \ .
\label{eq:adfl}
\end{equation}
Here we choose as  reference the cross-sectional area  $A^{\rm flat}$, as it is common for corrugated surfaces where the real contact area is not easily known. 
We find that the adsorption increases dramatically with increasing amplitudes $a$ and changes little for increasing polymer reservoir packing fractions  (Fig.~\ref{fig:ads}a).

In order to evaluate the effect of the chosen reference area, we also calculated the adsorption  using as a reference the real contact area $A^{\rm corr}$, i. e.
\begin{equation}
\Gamma_{\rm corr}^{*}=\frac{N_{c}^{\rm corr}-N_{c}^{\rm bulk}}{A^{\rm corr}}-\Gamma_{\rm flat} \ .
\label{eq:adcr}
\end{equation}
The increase in colloid adsorption for increasing amplitudes $a$ is now less dramatic  (Fig.~\ref{fig:ads}b), note the different y-axes scale) but  still clearly visible. 
Therefore, the increased surface area of the sinusoidal wall is not the only factor responsible for the increased adsorption and consequently for the stronger capillary condensation.

\begin{figure}
\includegraphics[width=8cm]{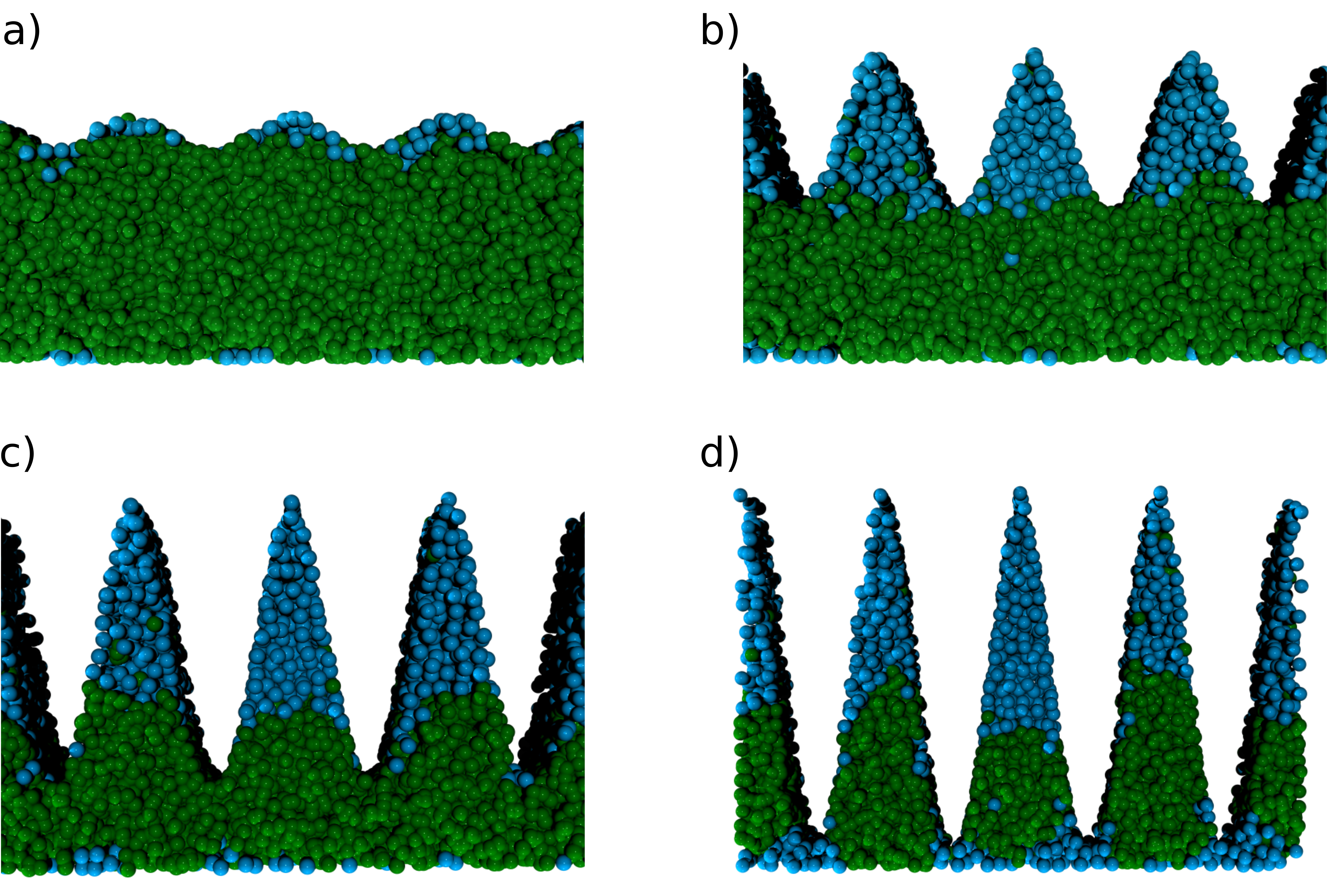}
\caption{Snapshots of the simulation at gas-liquid coexistence for the packing fraction corresponding to the equilibrium gas phase and polymer reservoir packing fraction $\eta_p^r=1.4$. The  colloidal particles (cyan/light grey spheres)  and the polymers (green/dark grey spheres) are confined between one flat wall (bottom) and one sinusoidal wall (top). The walls occupy the white sections of the figures. a) Amplitude $a/\sigma_c=1$ b) Amplitude $a/\sigma_c=5$ c) Amplitude $a/\sigma_c=9$ d) Amplitude $a/\sigma_c=13$}
\label{fig:snp}
\end{figure}

Interestingly, the simulation snapshots for state points at gas coexistence for polymer reservoir packing fraction $\eta_p^r=1.4$ (shown in Fig.~\ref{fig:snp}) show that colloids  are adsorbed inside the grooves of the sinusoidal walls, that is in the concave parts of the corrugated wall.

\subsection{Free energy curves}

\begin{figure}
\includegraphics[width=8.5cm]{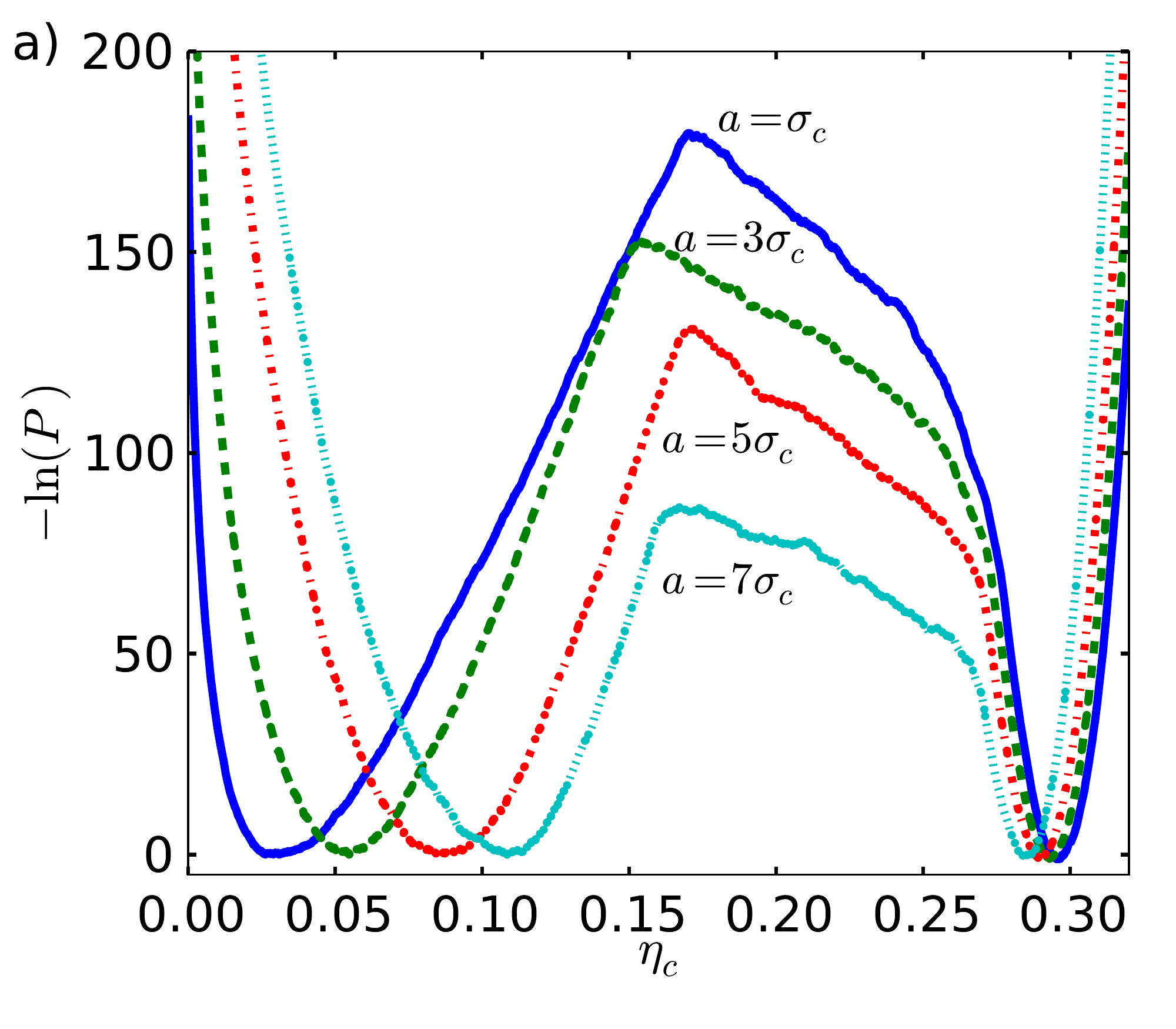}
\includegraphics[width=8.5cm]{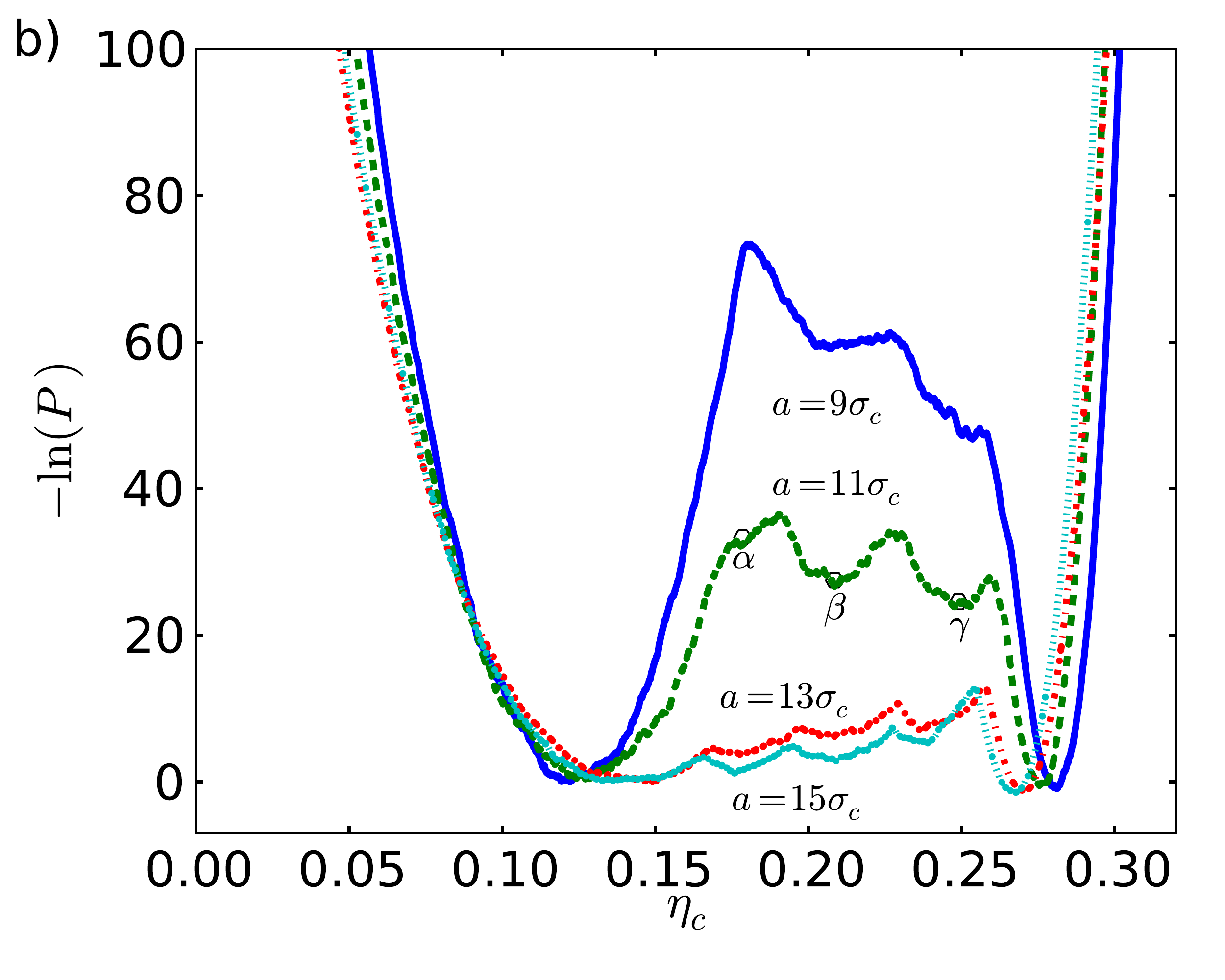}
\caption{Logarithm of the probability distribution (free energy) at polymer reservoir packing fraction $\eta_p^r=1.4$ and colloid chemical potentials at coexistence.  a) For amplitudes $a$=1, 3, 5, 7 $\sigma_{c}$. b) For amplitudes $a$=9, 11, 13, 15 $\sigma_{c}$. }
\label{fig:fen}
\end{figure}

\begin{figure}
\includegraphics[width=8.5cm]{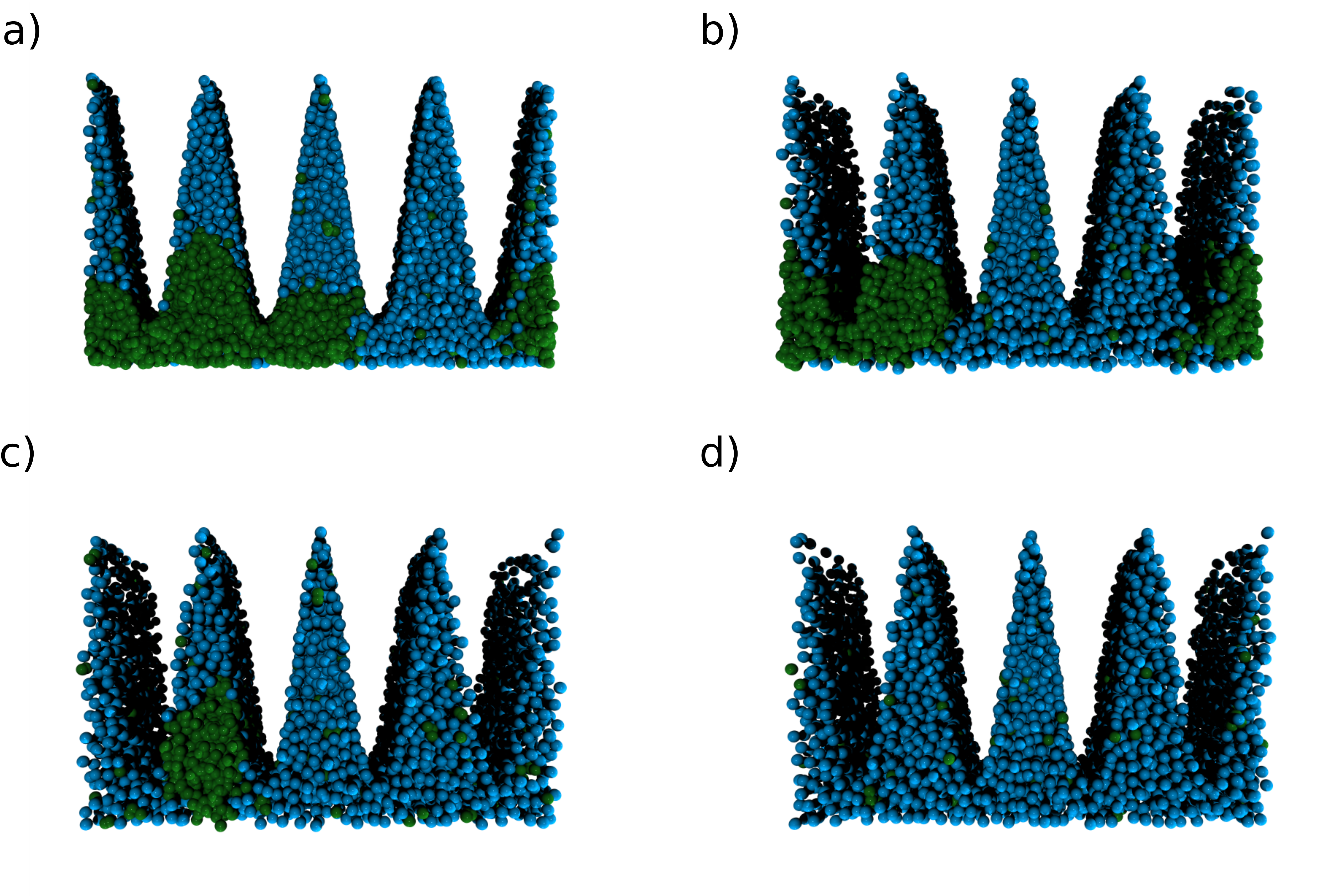}
\caption{Snapshots of the simulation at polymer reservoir packing fraction $\eta_p^r=1.4$ and amplitude $a/\sigma_c=11$. The  colloidal particles (cyan/light grey spheres)  and the polymers (green/dark grey spheres) are confined between one flat wall (bottom) and one sinusoidal wall (top). The walls occupy the white sections of the figures.   a) One channel filled with colloids at state point $\alpha$ in Fig.~\ref{fig:fen}b).   b)  Two  channels filled with colloids at state point $\beta$ in Fig.~\ref{fig:fen}b).  c) Three  channels filled with colloids at state point $\gamma$ in Fig.~\ref{fig:fen}b).  d) Equilibrium liquid phase. }
\label{fig:snp2}
\end{figure}

We next explore the dimensional crossover that occurs at amplitudes $a/\sigma_{c}$ comparable to the wall distance $H$. Consider  that the free energy (grand potential) difference with respect to a reference state is given by $-\ln(P(\eta_{c})|_{ \mu_c, \mu_p})$, where $P(\eta_{c})|_{ \mu_c, \mu_p})$ is the probability of observing the system at colloid packing fraction $\eta_{c}$.  
In Fig.~\ref{fig:fen}a) we plot the free energy curves for amplitudes $a$=1, 3, 5, 7 $\sigma_{c}$. The two minima correspond to the gas and liquid coexisting densities. For intermediate packing fractions $\eta_{c}$ the free energy decreases for increasing packing fraction, as expected for systems in slit geometry.
In Fig.~\ref{fig:fen}b) the free energy curves for amplitudes $a$=9, 11, 13, 15 $\sigma_{c}$ are shown. The approaching dimensional crossover is characterised by the appearance of a sequence of minima in the free energy curves. In our system the local minima appear first for $a/\sigma_c=9$ and becomes stronger and stronger for increasing amplitudes.  
These minima indicate the presence of metastable phases characterised by coexistence of channels filled with fluid phase and channels containing the gas phase.  
Each time a channel is completely filled we find an energy minimum due to the decreased  interfacial energy as shown in Fig.~\ref{fig:snp2} for polymer reservoir packing fraction $\eta_p^r=1.4$ and corrugation amplitude $a/\sigma_c=11$.
These phases are similar to the \emph{zebra phase} predicted for optically confined mixtures~\cite{Gotze2003,Vink:2011jl,Vink:2012dr}.

\section{Summary and conclusions} 
\label{conc}

We traced the phase diagram of model colloid-polymer mixtures confined between one flat wall and one corrugated wall with grand-canonical MC  simulations.
The corrugation was modelled  by a sinusoidal function of amplitude $a$. We found that for increasing values of $a$ the capillary condensation get stronger, i.e., the region of parameter space where the capillary condensation occurs becomes larger. 
We derive a Kelvin equation for the system that predicts that  capillary condensation in the system  is controlled by the Wenzel ratio $ R=\frac{A_{\rm corr}}{A_{\rm flat}}$  between areas of the corrugated and flat walls. 
We find very good agreement between simulation results and the Kelvin equation  prediction, indicating  that the increased contact area between the fluid and the substrate (due to corrugation) is the primary cause for the stronger capillary condensation. 
The analysis of the adsorption of colloidal particles shows a  strong preference of the colloids to adsorb at the corrugated wall in agreement with the stronger capillary condensation. The finding is also  corroborated by a visual inspection of the simulation snapshots, which show  a strong adsorption of  colloids in the grooves of the corrugated walls.  Nevertheless, we find an increase of adsorption for increased surface area also when the real contact area is used as reference. One possible explanation for this effect is that the depletion effect responsible for the colloid-wall effective attractive interaction is curvature dependent. The overlap area between  the particle and wall excluded volumes is different for different wall curvatures. Therefore the wall is more attractive at the concave side and less attractive at the convex edge. In our simulation we used a fixed wavelength and changing amplitudes, i.e., the  curvature of the corrugated walls would be different for the different amplitudes leading to a different effective wall-colloid attractive interaction. 
We also observe a dimensional crossover from a quasi-2D system to a quasi-1D system. The quasi-1D system occurs when the amplitude of the corrugated wall is equal to the average separation distance, i.e., all channels are decoupled. 
The free energy curves show that the approaching crossover is characterised by the appearance of a metastable phase, with partially filled channels, similar to the 'zebra' phase  predicted for optically confined colloid-polymer mixtures~\cite{Gotze2003,Vink:2011jl}.

In this paper we only considered capillary condensation and neglected two other relevant phenomena, namely wetting~\cite{Bonn:2009ha} and wedge filling transitions~\cite{Hauge:1992ct,Rejmer:1999hh,Parry:2000wk}. Given the finite system size as well as the finite wedge geometry our assumption is completely reasonable, but  an extension of this work would be to analyse the wetting and wedge filling behaviour  of a single  corrugated wall and compare to works on wedge filling transition in finite geometries~\cite{Tasinkevych:2006ep,Tasinkevych:2007fu,Parry:2007fz}.

A recent paper~\cite{Vink:2012dr} on fluids in periodic confinement showed unexpected correlations between the channels that deserve further investigation in the current system.

\providecommand*{\mcitethebibliography}{\thebibliography}
\csname @ifundefined\endcsname{endmcitethebibliography}
{\let\endmcitethebibliography\endthebibliography}{}

\end{document}